\documentclass[aps, prd, 12pt, amsmath, amssymb, superscriptaddress, tightenlines, nofootinbib, letterpaper]{revtex4-1}
\usepackage{amsmath}
\usepackage{comment}
\usepackage[pdftex]{graphicx}
\usepackage[pdftex, pdfstartview={FitH}, pdfnewwindow=true, colorlinks=true, citecolor=blue, filecolor=blue, linkcolor=blue, urlcolor=blue, pdfpagemode=UseNone, bookmarks=false]{hyperref}
\usepackage{float}
\usepackage{subcaption}
\usepackage{caption}
\usepackage{xcolor}
\usepackage{soul,ulem}

\usepackage{orcidlink}

\newcommand{\eps}{\ensuremath{\epsilon} }

\newcommand{\fig}[1]{Fig.~\ref{#1}}

\newcommand{\psib}{\overline{\psi}}

\newcommand{\chib}{\overline{\chi}}
\newcommand{\tr}{\mathrm{tr}}

\newcommand{\gwen}[1]{\textcolor{black}{#1}}

\begin{document}
\title{Searching for symmetric mass generation with staggered fermions in four dimensions}
\author{Nouman Butt}
\affiliation{Department of Physics, University of Rhode Island, Kingston, RI 02881, United States}
\author{Simon Catterall}
\affiliation{Department of Physics, Syracuse University, Syracuse, NY 13244, United States}
\author{Gwen Hartshaw}
\affiliation{Department of Physics, Syracuse University, Syracuse, NY 13244, United States}
\thanks{gehartsh@syr.edu}
\author{Anna Hasenfratz}
\affiliation{Department of Physics, University of Colorado, Boulder, CO 80309, United States}

\begin{abstract}
    {We conduct numerical simulations to map out the phase diagram and critical behavior of a lattice
    Higgs model composed of two massless staggered fermion fields forming a doublet under a global $SU(2)$ and coupled to a scalar field
    in the adjoint representation of the group. The scalar action consists of a potential comprising quadratic and quartic terms and a scalar kinetic term. At fixed quartic coupling we 
    explore a two-dimensional parameter space finding a
    massless symmetric phase at weak coupling and a massive symmetric
    phase (SMG phase) at strong coupling. An intermediate anti-ferromagnetic phase separates these two regimes. These results are consistent with leading order weak and strong coupling expansions. We
    find that the critical lines bounding the intermediate phase merge at a unique point where all fermion bilinear condensates vanish but fermion susceptibilities diverge as non-trivial powers of the
    lattice size. We conjecture that this merged point corresponds to a multicritical point
    and may describe a phase consisting of a condensate of certain topological defects.
    }
\end{abstract}
\maketitle

\newpage
\section{Introduction}

The last decade has witnessed a growing interest and exploration of new models
of strongly coupled fermions in both the condensed matter and particle physics communities \cite{Fidkowski:2009dba,You:2014oaa,You:2014vea,Ayyar:2014eua,Ayyar:2015azp,Slagle_2015,Catterall:2015zua,Catterall:2016dzf,Butt:2018nkn,He:2016ixs,Wang:2018rjg,Wu:2019eag,Razamat:2020kyb,Tong:2021phe,CresswellHogg:2024fmg,Butt:2024kxi}. One feature of
particular interest has been the possible generation of mass without explicit or spontaneous symmetry breaking.
In low dimensions there is considerable evidence in favor of such SMG (symmetric mass generation) phases \cite{Fidkowski:2009dba}, but the situation in four
dimensions is less clear. 
There is a long history of lattice studies of strongly coupled scalar-fermion systems motivated by the possibility of nonperturbative fermion mass generation. Early work on scalar-fermion models found rich phase diagrams containing para-, ferro-, antiferro- and ferrimagnetic phases and even a strongly coupled phase with SMG-like mass generation \cite{Hasenfratz:1988vc,Hasenfratz:1989jr,Bock:1990tv,Lee:1989xq,Lee:1989mi}. However, no evidence was found for non-trivial continuous phase transitions where the SMG phase could survive into 
the continuum.  These studies already emphasized that strong Yukawa interactions can induce unusual phases, but they were not designed to realize an SMG phase protected by the anomaly structure of the staggered-fermion symmetries, and the interpretation in terms of symmetric mass generation without fermion bilinear condensation was not the focus. The present work revisits this class of questions in a model whose lattice symmetries and representation content are chosen so that a symmetric massive phase is allowed.

Ref \cite{Butt:2021koj} studied a theory of four reduced staggered fermions gauged under an $SU(2)\times SU(2)$ symmetry and provided evidence in favor of an SMG phase driven by confinement. Unlike QCD-like theories a four fermion rather than bilinear condensate was observed to form. However this
study used small lattices and no attempt was made to take a continuum limit. Ref \cite{Butt:2024kxi} studied a theory of staggered fermions with an $SU(2)$ gauge symmetry and found evidence for an unusual phase structure and 
a new strongly coupled phase which was conjectured to be an SMG phase. In the current work we have attempted
to follow up on an earlier work \cite{Butt:2018nkn} which attempted to find such a phase in a Higgs-Yukawa model using {\it reduced}  staggered fermions
without gauge interactions. In our new work we have mapped this latter model
into a two flavor staggered model, expanded the parameter space and attempted a much more exhaustive
exploration of the phase diagram. We have also identified an interesting feature of the model - in the continuum
it admits an unusual type of topological field configuration - the Hopf defect - that can play a role analogous
to that of vortices in the 2d XY model and disorder the vacuum without breaking symmetries \cite{He:2014gqa,Bruckmann:2000xd}.

\section{Lattice model and symmetries}
The action we consider takes the form
\begin{equation}
\begin{aligned}
    S = &\sum_x\bar\psi \left( \eta\cdot\Delta+iy\sigma\right) \psi + \frac{1}{2}\sum_x |\sigma|^2 
     -\frac{\kappa}{2}\sum_{x} \sigma\Box\sigma + \frac{\lambda}{4}\sum_x|\sigma|^4
\end{aligned}
\label{origaction}
\end{equation}
where $\psi$ is a two component staggered fermion field. The difference operator is $\Delta_\mu\psi = \psi(x+\mu) -\psi(x-\mu)$ while $\eta^\mu(x) = (-1)^{\sum_{i=1}^{\mu-1} x_i}$ is the usual staggered fermion phase. The field $\sigma$ is a scalar in the adjoint representation of a global SU(2) symmetry  whose norm is defined by $|\sigma|^2=\tr(\sigma^\dagger\sigma)$ and 
$$\Box\sigma = \Delta_\mu^2\sigma = \sum_\mu \sigma(x+2\mu)+\sigma(x-2\mu) - 2\sigma(x)$$ 
is the discrete Laplacian acting on the {\it block} lattice \gwen{which is the hypercubic grid with effective lattice spacing twice that of the fundamental (``{\it fine}'') lattice}.\footnote{\gwen{This choice of Laplacian therefore couples sites of the same relative position in each hypercube (and thus having the same lattice parity) while the traditional lattice Laplacian $\Delta^+_\mu\Delta^-_\mu$ (acting on the fine lattice) would instead couple nearest-neighbor sites (of different parity). See section~\ref{deriv-exp} for more on why this Laplacian operator was chosen.}}

The action in Eq.~(1) is similar to the scalar-fermion lattice models studied in the early Higgs-Yukawa literature, but differs in several important respects. The older models employed staggered fermions coupled to scalar fields with the usual nearest-neighbor hopping term, and their phase diagrams were organized in terms of paramagnetic, ferromagnetic, antiferromagnetic and ferrimagnetic order. In contrast, the present theory uses massless staggered fermions coupled to a scalar via an $SO(4)$ invariant Yukawa interaction whose structure
is chosen to favor a four fermion condensate at strong coupling. Crucially the symmetries
of the model include both shift symmetries and a discrete $Z_4$ symmetry. In addition, we use a block-lattice Laplacian for the scalar kinetic term. As discussed below, this derivative operator is natural when the effective action is expanded about an antiferromagnetic background which is the favored broken symmetry phase of the model.
This field content leads to a simpler phase diagram with just free, antiferromagnetic and SMG like phases.

This fermionic sector can be mapped into the model with four reduced staggered fermions that was studied in \cite{Butt:2018nkn} - see appendix~\ref{map} for details.
In the current study we have enlarged the parameter space by adding a quartic term and replacing the original scalar hopping term by the block Laplacian. In all the work presented in this paper we set $\lambda=1.0$.
As detailed in appendix~\ref{map}, the kinetic term is actually invariant under an
$U(4)$ global symmetry which is broken to $SO(4)\times Z_4$ by the Yukawa term. The $Z_4$ symmetry is just a remnant of the usual $U_\epsilon(1)$ symmetry and contains
the elements $\omega\in \left[1,i\epsilon(x),-1,-i\epsilon(x)\right]$ 
\begin{align}
    \psi(x)&\to \omega\psi(x)\nonumber\\
    \psib(x)&\to \omega\psib(x)\nonumber\\
    \sigma(x)&\to \omega^{2}\sigma(x)
\end{align}
\gwen{where $\epsilon(x)=(-1)^{\sum_{i=1}^4 x_i}$ is the parity of site $x$.}
The action is also invariant under the shift symmetry:
\begin{align}
    \psi(x)&\to \xi_\nu(x)\psi(x+\nu)\nonumber\\
    \psib(x)&\to \xi_\nu(x)\psib(x+\nu)\nonumber\\
    \sigma(x)&\to -\sigma(x+\nu)
\end{align}
\gwen{where $\xi_\nu(x)=(-1)^{\sum_{i=\nu+1}^4 x_i}$ is the usual shift-symmetry phase factor.}
The spin-$Z_4$ symmetry crucial for SMG is related to a combination of the onsite $Z_4$ symmetry
and the hypercube shift which is obtained by combining elementary shifts along the four orthogonal
directions 
\begin{equation}
    \chi(x)\to  i\eps(x)\xi_1(x)\xi_2(x+\hat{1})\xi_3(x+\hat{1}+\hat{2})\xi_4(x+\hat{1}+\hat{2}+\hat{3})\chi(x+\hat{1}+\hat{2}+\hat{3}+\hat{4})
\end{equation}
In the naive continuum limit this behaves as a discrete axial rotation on the Dirac fermions that are built from the staggered fields.

After integration over the fermions the scalar effective action will remain
invariant under all these symmetries. However spontaneous breaking of symmetries can still
occur and would be signaled by a non-zero expectation value for an appropriate order parameter.
For example, the fermion bilinear 
$\bar{\psi}^a\psi^a$ is invariant under the $SO(4)$ and shift symmetries but breaks
the $Z_4$ symmetry. A necessary condition to avoid such spontaneous breaking and thereby achieve
an SMG phase is that all 't Hooft anomalies for the global symmetries must vanish. To look
for such 't Hooft
anomalies one can look for obstructions to gauging the symmetry.
There is clearly no barrier to gauging the lattice action under $Z_4$ - one merely has to
insert appropriate $Z_4$-valued gauge links into the kinetic term. But it is not hard to
see that the fermion measure is not invariant under {\it local} $Z_4$ rotations unless the number of
staggered fields is even. This is clearly satisfied for our model so
there is therefore no 't Hooft anomaly associated to the $Z_4$ symmetry. In addition one cannot have anomalies for orthogonal symmetries such as $SO(4)$ \gwen{because such symmetries only have real or pseudoreal representations which can never be anomalous}. It is harder to analyze possible 't Hooft anomalies for the
shift symmetries because they are not onsite - they involve lattice translation. However we will show numerical results later that are consistent with a lack of spontaneous symmetry breaking
for these symmetries too.~\footnote{The model has an
additional symmetry - axis inversion (reflection) which also suffers from potential 't Hooft anomalies but in fact this anomaly is also canceled for even numbers of staggered fields.}. Of course, anomaly
cancelation is a necessary but not sufficient condition to realize an SMG phase - the dynamics of the model is also important. Indeed,
we will argue
that this model possesses additional interesting structure and dynamics
which may play a role in 
realizing an SMG phase that could survive into the continuum.

On integration over the fermions one
obtains ${\rm det}(\eta \cdot\Delta +iy\sigma)$ where the fermion operator is an anti-hermitian matrix. This fact together with the $SU(2)$ symmetry is sufficient to show that the determinant is real, positive and can be
simulated using an RHMC algorithm - see appendix~\ref{sign}.

\section{Weak and strong coupling expansions }
It is helpful to perform a weak coupling expansion on the fermion determinant which we
write as 
\begin{equation}-{\rm Tr}\log{(\eta\cdot \Delta +iy\sigma)}=\sum_{n=1} \left(-1\right)^{n}{\rm Tr}(G(x,y)iy\sigma(y))^n
\end{equation}
where $G=(\eta \cdot\Delta)^{-1}=\frac{\eta \cdot\Delta}{\Delta^2}$.
The first order term vanishes~\footnote{In fact all odd order terms vanish because of the $Z_4$ symmetry.} while at second order one finds
\begin{equation}S_{\rm eff}=\frac{y^2}{2}\sum_{x,y} |G(x,y)|^2\sigma^a(x)\sigma^a(y)\end{equation}
where we have employed $G(x,y)=-G(y,x)$. Clearly this gives
an antiferromagnetic nearest neighbor interaction. Since $G^2(x,y)\sim \frac{1}{|x-y|^6}$ the nearest
neighbor term dominates in an expansion in $|x-y|$ and we deduce that the dynamics of the model
favors
the formation of an antiferromagnetic ground state. Indeed, if we compute the one loop effective potential for
$\sigma$ by expanding around a constant
antiferromagnetic background $\sigma(x)=\mu\epsilon(x) \tau^3$ one finds
\begin{align}
    V_{\rm eff}(\mu)&=-\frac{1}{2}{\rm Tr}\;\log{\left(\left[\eta \cdot\Delta+iy\mu\epsilon\tau^3\right]\left[-\eta\cdot\Delta-iy\mu\epsilon\tau^3\right]\right)}\nonumber\\
    &=-\frac{1}{2}{\rm Tr}\;\log{\left(-\Delta_\mu^2+y^2\mu^2\right)}
    \label{Anti}
    \end{align} 
where the linear term in $y$ vanishes because $\epsilon$ anti-commutes with $\Delta_\mu$. Subtracting the value at $y=0$ and expanding the log we find
\begin{equation}
    V_{\rm eff}(\mu)=-\frac{y^2\mu^2}{2}{\rm Tr}\;\left(\frac{1}{-\Delta_\mu^2}\right)+\ldots\end{equation}
Adding the classical potential the total effective potential is given by
\begin{equation}V_{\rm eff}(\mu)= \frac{1}{2}(1-\alpha y^2)\mu^2+\frac{\lambda}{2}\mu^4\end{equation}
Clearly loop effects cause the coefficient of $\mu^2$ to change sign for some $y>y_c$ giving 
rise to a non-zero vacuum
value for $\mu$ and antiferromagnetic order for at least some range of $y$.

To complement this weak coupling calculation we can also perform an expansion in $1/y$. Including terms to
second order we find 
\begin{equation}S_{\rm eff}=-{\rm Tr}\log{\big[\left(I+\frac{1}{iy\mu^2}\eta\cdot\Delta \sigma\right)iy\sigma\big]}=-\sum_x \log\sigma^2(x)+\frac{1}{2y^2\mu^4}\sum_{x,y}|G^{-1}(x,y)|^2\sigma^a(x)\sigma^a(y)\end{equation}
where $G^{-1}=\eta \cdot\Delta$ and we have used $\sigma^{-1}=\frac{1}{\mu^2}\sigma$ as appropriate to
a phase where $\sigma^2=\mu^2$. 
The leading term as $y\to\infty$ 
favors a non-zero value for $\sigma^2=\mu^2$ but allows the direction in field space to vary randomly from
one site to another so there is no spontaneous symmetry breaking. It corresponds to a lattice SMG phase.
The correction term is a local antiferromagnetic coupling 
\begin{equation}
    \frac{1}{y^2\mu^4}\sum_{x,\mu} \left[\sigma^a(x)\sigma^a(x+\mu)+\sigma^a(x)\sigma^a(x-\mu)\right]
\end{equation}
Taking these analytic results together 
we thus expect a symmetric weak coupling phase at small $y$, an antiferromagnetic phase at intermediate $y$
and eventually a symmetric strong coupling phase for $y\to\infty$. 

\section{A derivative expansion}
\label{deriv-exp}

It is interesting to consider what other terms can arise in the effective action for $\sigma$ if we
expand the fluctuations in $\sigma(x)$ around an antiferromagnetic background 
by writing
$\sigma(x)=\mu\epsilon(x)n^a(x)\tau^a$ where $n^a(x)n^a(x)=1$. The effective action that generalizes
eqn.~\ref{Anti} now reads
\begin{equation}S_{\rm eff}=-\frac{1}{2}{\rm Tr}\;\ln{\left(-\Box+M^2+iM\epsilon(x)[\eta_\mu(x)\Delta_\mu, n(x)]\right)}\end{equation}
where $M=y\mu$.
We can simplify the term linear in $M$ by keeping track of how it acts on a slowly varying
test function $f(x)$: 
\begin{align}
&iM\epsilon(x)\eta_\mu(x)[n(x+\mu)f(x+\mu)-n(x-\mu)f(x-\mu)-n(x)\Delta_\mu f(x)]\nonumber\\
&\sim iM\epsilon(x)\eta_\mu(x)f(x)\Delta_\mu n(x)
\end{align}
where the last line is the leading $O(a)$ approximation.
Thus the part of the fermion matrix that depends linearly on $M$ can be written
\begin{equation}iM\epsilon(x)\eta_\mu(x)\Delta_\mu n^a(x)\tau^a\end{equation}
Again subtracting the
value at $y=0$ we find the effective action can be written
\begin{align}S_{\rm eff}&=-\frac{1}{2}{\rm Tr}\;\ln{\left[\left(\frac{-\Box+M^2}{-\Box}\right)\left(I+\frac{iM\tau^a\epsilon(x)\eta_\mu(x)\Delta_\mu n^a(x)}{-\Box+M^2}\right)\right]}
\end{align}
The first factor inside the logarithm just gives us the effective potential again but the second factor gives
us something new:
\begin{align}
S_{\rm eff}&=-\frac{1}{2}{\rm Tr}\,\ln{\left[I+\frac{i\tau^a\epsilon(x)\eta_\mu(x)\frac{\Delta_\mu}{M} n^a(x)}{-\frac{\Box}{M^2}+1}\right]}\nonumber\\
&=-\frac{1}{2}{\rm Tr}\,\ln{\left[I+\left(1+\frac{\Box}{M^2}+..\right)i\tau^a\epsilon(x)\eta_\mu(x)\frac{\Delta_\mu}{M} n^a(x)\right]}
\end{align}
We now perform a derivative expansion in powers of $\frac{\Delta_\mu}{M}$. The leading term is quadratic in
derivatives 
\begin{equation}
    -\frac{1}{4M^2}{\rm Tr}\left(\epsilon(x)\eta_\mu(x) \Delta_\mu n^a(x)\tau^a\right)^2=
    -\frac{1}{4M^2}\sum_x n^a(x) \Box n^a(x)
    \label{quad}
\end{equation}
where $\Box$ is the block lattice Laplacian $\Box=\Delta_\mu^2$.
It is a generic
non-linear sigma model term that would arise in the effective action description of fluctuations
around a broken phase for any number of staggered
fermion fields. Notice that the term
\begin{equation}
  -\sum_x  \sigma(x)\Box\sigma(x)
\end{equation} couples fields with the same site parity and
vanishes on an antiferromagnetic background.  It hence naturally describes the action cost of fluctuations
around such a ground state unlike the usual lattice Laplacian $\Delta_\mu^+\Delta_\mu^-$ which couples nearest neighbors and only vanishes for ferromagnetic backgrounds. It is this observation that lies behind the
choice of this operator in our bare lattice action given in eqn.~\ref{origaction}.

At the next non-zero order in the expansion there are two terms $S=S_1+S_2$ where
\begin{align}
    S_1&=\frac{1}{2M^4}{\rm Tr}\,(n^a(x)\Box^2n^a(x))=\frac{1}{2M^2}\sum_x \sigma(x)\Box^2\sigma(x)\nonumber\\
    S_{2}&=\frac{1}{16M^4}{\rm Tr}\,(\epsilon^{abc}\eta_\mu(x)\Delta_\mu(x,y) n^a(y)\eta_\nu(x)\Delta_\nu(x,z) n^b(z))^2
\end{align}
where, for clarity, in the second term we have written out the spacetime indices on the symmetric difference
explicitly. $S_1$ gives a higher order correction to the quadratic action for $\sigma$. However, $S_2$ is a new quartic operator whose structure depends on the fact that the scalars live in the adjoint representation of $SU(2)$. Notice that the antisymmetry in the group indices is consistent with the properties of the staggered phases
$\eta_\mu(x)\eta_\nu(x+\mu)+\eta_\nu(x)\eta_\mu(x+\nu)=\delta_{\mu\nu}$ which in turn reflects the anti-commutation properties of the Dirac gamma matrices in the continuum. In the continuum $S_2$ takes the form of a Skyrme term \cite{Faddeev:1996zj}.
\begin{equation}(\partial_\mu n\times \partial_\nu n)^2\end{equation}
This term is a marginal operator in four dimensions and
its presence allows for the possibility of topological defects which we will describe in the
next section.
\begin{figure}[h]
    \centering
    \begin{subfigure}[b]{0.48\linewidth}
        \centering
        \includegraphics[width=\linewidth]{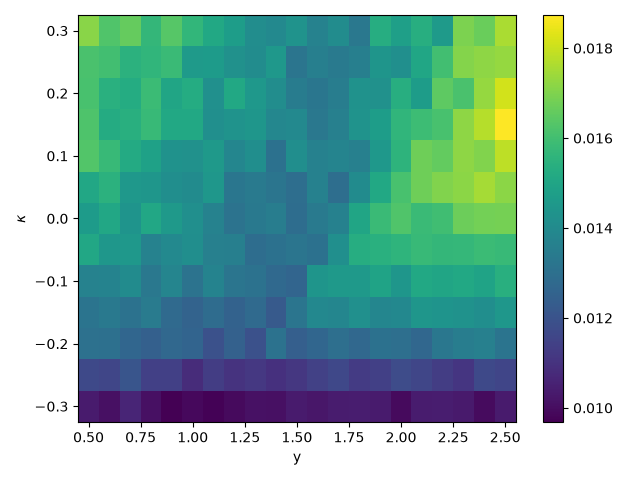}
        \caption{$<\Sigma>$ at $\lambda=1.0$}
    \end{subfigure}
    \begin{subfigure}[b]{0.48\linewidth}
        \centering
        \includegraphics[width=\linewidth]{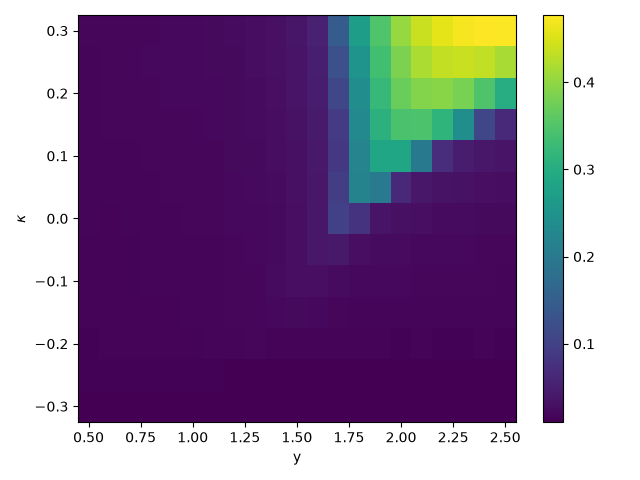}
        \caption{$<\Sigma_{\rm stag}>$ at $\lambda=1.0$}
    \end{subfigure}
    \begin{subfigure}[b]{0.48\linewidth}
        \centering
        \includegraphics[width=\linewidth]{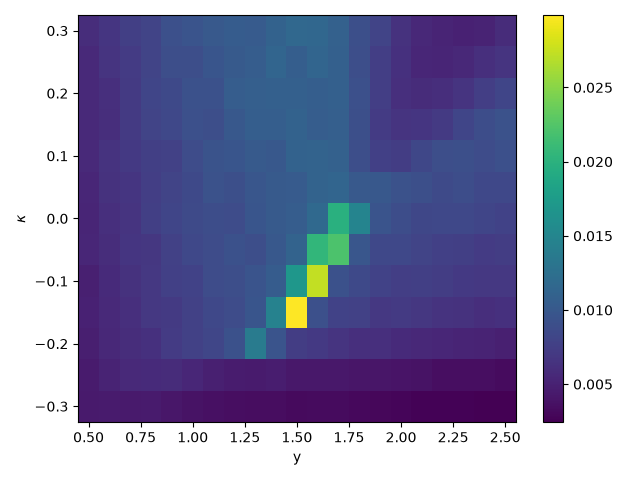}
        \caption{$<\phi>$ at $\lambda=1.0$}
    \end{subfigure}
    \begin{subfigure}[b]{0.46\linewidth}
        \centering
        \includegraphics[width=\linewidth]{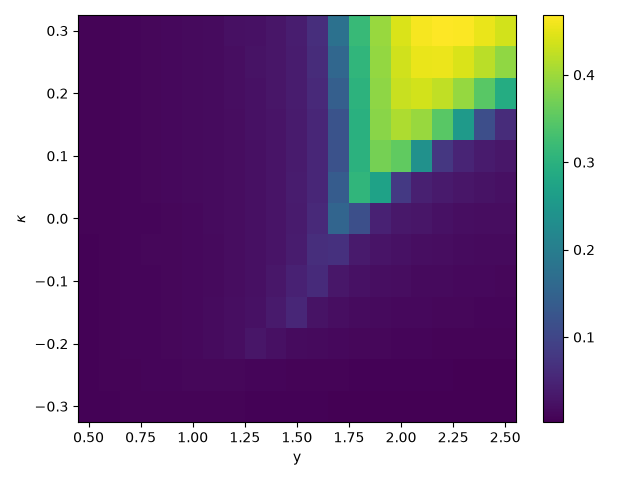}
        \caption{$<\phi_{\rm stag}>$ at $\lambda=1.0$}
    \end{subfigure}
    \caption{Heatmaps for $8^4$ lattice at $\lambda=1.0$}
    \label{heat}
\end{figure}
\section{Topological defects}

This model admits topologically non-trivial field configurations called Hopf defects in the continuum.~\footnote{There is another class of possible defects corresponding to
a non-trivial map between the theory compactified on the four sphere and the group manifold of $SU(2)$ and classified by $\Pi_4(S^3)=Z_2$. We thank Cenke Xu for pointing this out.} The topological character is connected to a non-trivial map $\Pi_3(S^2)$ between
the $S^2$ vacuum manifold arising from the constraint $n^a n^a=1$ and the $S^3$ spacetime
boundary at infinity. 
These Hopf defects can be constructed explicitly by 
first changing variables from $n^a$ to $SU(2)$ matrices $P$ via
\[n^a(x)\tau^a=P(x)\tau_3P^\dagger(x)\]
The constraint $n^an^a=1$ is now encoded in the $SU(2)$ character of $P$. Actually the $n^a$ are unchanged under the local phase
change $P(x)\to e^{i\alpha(x)\tau_3}P(x)$. This is needed to reduce
the three dof in a SU(2) matrix to two to match the two independent $n^a$'s. 
If we parametrize the $P$ field as
\[\left(\begin{array}{cc}\alpha_1+i\alpha_2&-\alpha_3+i\alpha_4\\\alpha_3+i\alpha_4&\alpha_1-i\alpha_2\end{array}\right)\] 
with $\sum_{i=1}^4\alpha_i^2=1$ the Hopf defect corresponds to the choice
\[\alpha_i=\frac{x_i}{r}\]
Naively this choice corresponds to a map between the boundary sphere $S^3$ and an $S^3$ associated with the group manifold of $SU(2)$. Actually because the $P$ matrices are to be identified up to a local phase this actually corresponds to mappings between $S^3$ and $S^2$. 

In the absence of the quadratic term in eqn.~\ref{quad}
the action is dominated by the quartic Skyrme term and because this term is marginal the action of such Hopf
defects will diverge logarithmically with system size. Thus 
defects are naively suppressed in
the large volume limit. However, in the path integral one should integrate over all locations 
of such a defect and this leads to a logarithmic
contribution to the entropy of such configurations which can compete with their action. It is thus
possible that the resulting free energy minimum corresponds to a condensate of such defects in certain
regimes. Notice that such a condensate would break no symmetries and would hence be a candidate for a continuum
scalar field realization of an SMG phase.

This situation is analogous to the XY model
in two dimensions where a non-trivial map $\Pi_1(S^1)$ arises for vortex configurations
which wind around the $S^1$ vacuum manifold on the boundary circle at infinity. In
the XY case the action of such defects also diverges logarithmically with the system size. 
This logarithmic action cost for forming a vortex competes with a logarithmic
entropy associated with the choice of the coordinates of
the center of the vortex and induces a (BKT) phase
transition to a defect condensed phase at sufficiently high temperature \cite{Berezinskii:1971,Kosterlitz:1973xp}.

However there is one significant difference between
the XY and Hopf defect pictures. In the former the marginal operator that
generates the logarithmic action is the
leading term in the sigma model action. In our model it is subleading to a quadratic term that
represents the usual kinetic energy of the sigma model. To see
defect condensation one would then need to tune the bare sigma model
action to set the coupling of this term to zero in the I.R.
This provides the
rationale for our exploration of an expanded parameter space in the theory
which includes a appropriate quadratic kinetic term for the scalars. In other words we conjecture it may be
possible to tune $\kappa$ in such a way as generate a new fixed point in the lattice theory whose continuum description would include an SMG phase consisting of condensed Hopf defects.
\begin{figure}
\includegraphics[width=0.6\textwidth]{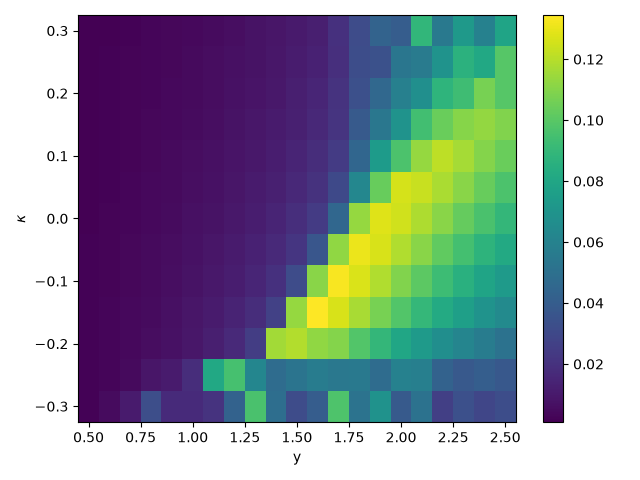}
\caption{Four fermion condensate $<O_4>$ at $\lambda=1.0$}
\label{4cond}
\end{figure}

\section{Phase Diagram}
Our initial goal was to map out the phase diagram of this model in the two dimensional parameter space $(\kappa,y)$ for fixed quartic coupling $\lambda=1.0$ using a modest lattice of size $8^4$.
Since we are interested in symmetric phases we have focused on measurements of fermion bilinears
which are $SO(4)$ and shift invariant
but break $Z_4$~\footnote{We will also show later
measurements of the one link bilinear which is $SO(4)$ and $Z_4$ invariant but breaks shift symmetry.}.
They are defined by 
\begin{equation}
    \phi = \frac{1}{V}\Big|\sum_x\bar\psi(x)\psi(x)\Big|\quad{\rm and} \quad \phi_{\rm stag}=\frac{1}{V}\Big|\sum_x\varepsilon(x)\bar\psi(x)\psi(x)\Big|
\end{equation}
Taking the absolute value of $\phi$ and $\phi_{\rm stag}$ before averaging over configurations allows us to look carefully for spontaneous symmetry breaking. In a situation where the symmetry is
unbroken these observables decrease with increasing lattice size $L$ while in the broken case they remain
$L$ independent at criticality.
We have also examined both ferro and anti-ferromagnetic order parameters of the scalar field given by
\begin{equation}
    \Sigma=\frac{1}{V}\Big|\sum_x \sigma(x)\Big|\quad{\rm and}\quad \Sigma_{\rm stag}=\frac{1}{V}\Big|\sum_x \varepsilon(x)\sigma(x)\Big|
\end{equation} 
We also measure $|\sigma^2|$ and a possible four fermion condensate $O_4=<\bar{\psi}^1\psi^1\bar{\psi}^2\psi^2>$.
Notice that these latter two observables are invariant under all lattice symmetries.
We have also measured two fermion susceptibilities defined by 
\begin{equation}
    \chi = V(<\phi^2>-<\phi>^2)\quad{\rm and}\quad \chi_{\rm stag} = V(<\phi_{\rm stag}^2>-<\phi_{\rm stag}>^2)
\end{equation}
where $<\cdot>$ denotes the average over configurations.

\begin{figure}
    \centering
    \includegraphics[width=0.6\linewidth]{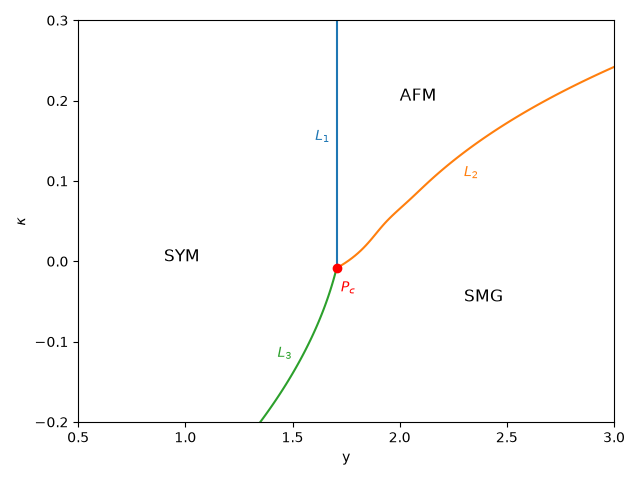}
    \caption{The phase diagram of the model. There is an unbroken symmetric (SYM) phase at small $y$ and an SMG phase at large $y$. Between them in the region above $\kappa_c$ is a broken antiferromagnetic (AFM) phase. These three phases meet at a point $P_c = (y_c,\kappa_c)$. At large negative $\kappa$ there is a frustrated phase whose boundary with the other phases we did not explore.}
   \label{fig:phase}
\end{figure}

We ran a series of simulations across a grid of values in the $(\kappa,y)$ space. 
For positive $\kappa$, the block kinetic term provides a positive stiffness for fluctuations about the antiferromagnetic background and therefore stabilizes the AFM phase. Reducing $\kappa$ weakens this stiffness, allowing the two AFM phase boundaries to approach and eventually merge. The heat maps 
in fig.~\ref{heat} show results for the ensemble average of the
corresponding bilinears over the entire grid.
Notice that the anti-ferromagnetic fermion bilinear $<\phi_{\rm stag}>$ and corresponding scalar $<\Sigma_{\rm stag}>$ are highly correlated as one
might expect and reveal a triangular phase of broken symmetry in the upper right hand region of the
phase diagram. The critical lines forming the left and right boundaries of this anti-ferromagnetic phase appear to
merge at a single point $P_c=(\kappa_c,y_c)$. There is some evidence that a single merged critical line
extends from this point to negative $\kappa$ as is visible in the heatmap
of the four fermion operator $O_4$ shown in figure~\ref{4cond}.
We can summarize our results in the phase diagram in fig.~\ref{fig:phase}. We now turn to a more detailed study both of the broken phase and the merged point $P_c$.

\section{Phase transitions for $\kappa\ge \kappa_c$}
\begin{figure}[h]
    \centering
    \begin{subfigure}[b]{0.48\linewidth}
        \centering
        \includegraphics[width=\linewidth]{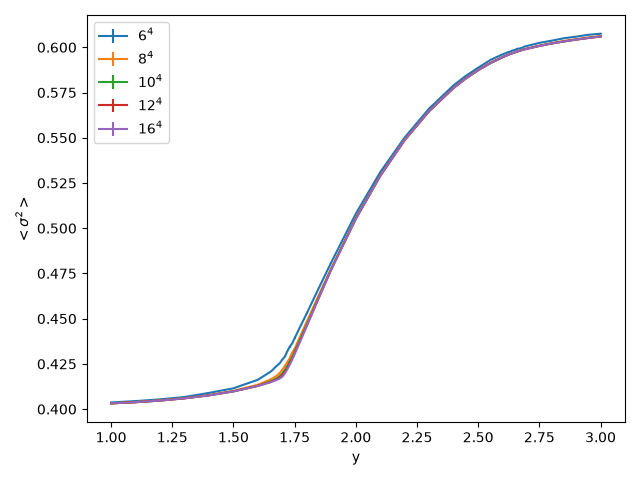}
        \caption{$<|\sigma^2|>$ at $\kappa=0.2$}
    \end{subfigure}
    \begin{subfigure}[b]{0.48\linewidth}
        \centering
        \includegraphics[width=\linewidth]{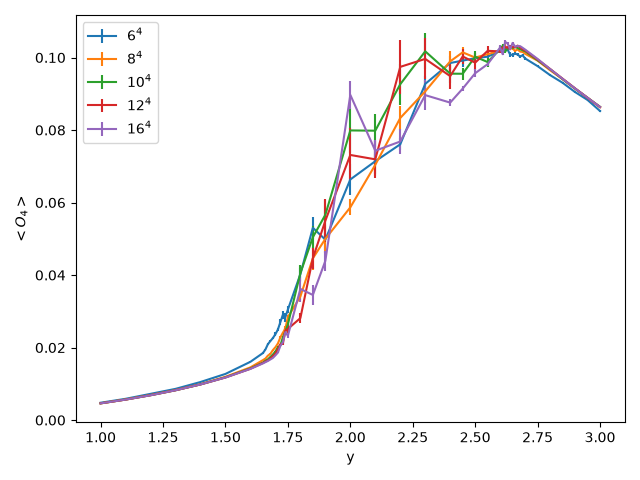}
        \caption{$<O_4>$ at $\kappa=0.2$}
    \end{subfigure}
\caption{$|\sigma^2|$ and four fermion condensate $O_4$ vs $y$ at $\kappa=0.2$}\label{sigma2}
\end{figure}

\begin{figure}\centering
    \begin{subfigure}[b]{0.48\linewidth}
        \centering
        \includegraphics[width=\linewidth]{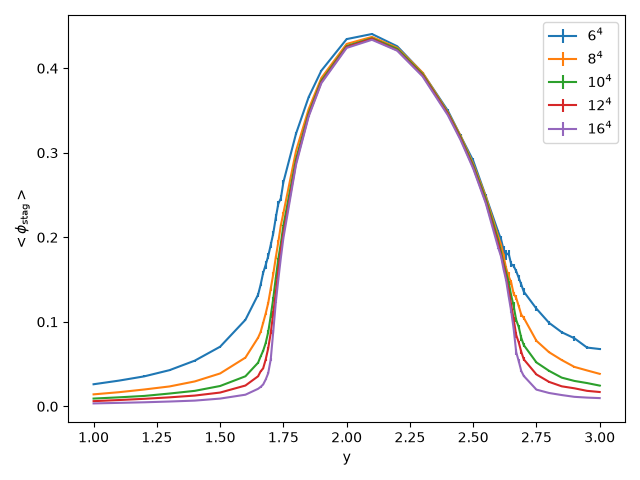}
        \caption{$<\phi_{\rm stag}>$ at $\kappa=0.2$}
    \end{subfigure}
    \begin{subfigure}[b]{0.48\linewidth}
        \centering
        \includegraphics[width=\linewidth]{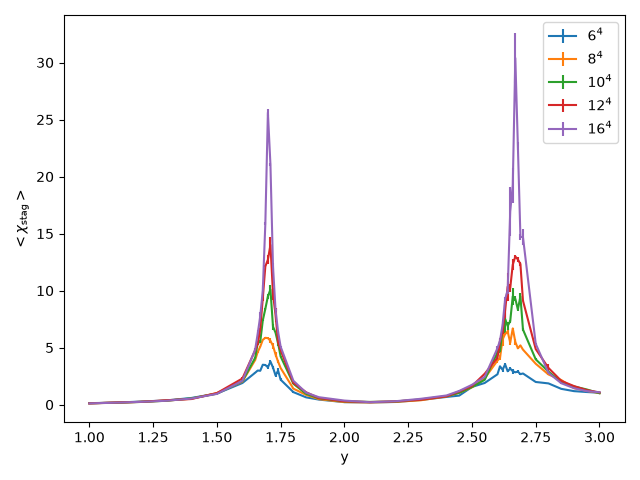}
        \caption{$<\chi_{\rm stag}>$ at $\kappa=0.2$}
    \end{subfigure}
    \caption{Anti-ferromagnetic bilinear and its susceptibility vs $y$ at $\kappa = 0.2$ and $\lambda=1.0$}
    \label{kappa0.2}
\end{figure}

\begin{figure}
    \centering
    \includegraphics[width=0.5\linewidth]{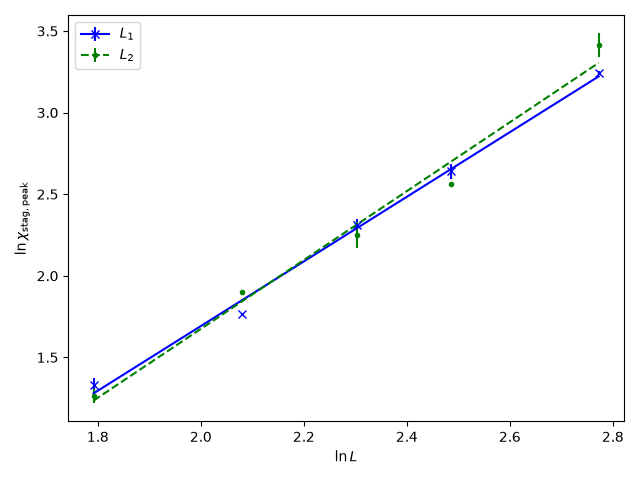}
    \caption{Peak $\chi_{\rm stag}$ versus $L$ for $\kappa=0.2$ and $\lambda = 1.0$ along the phase transition boundaries $L_1$ and $L_2$.}
    \label{loglog}
\end{figure}

We first focus on the region that contains the anti-ferromagnetic phase.
In fig.~\ref{sigma2} we show plots of the $|\sigma^2|$ and the four fermion operator $O_4$ at $\kappa=0.2$ as a function of the
Yukawa coupling $y$. There is clear evidence of a weak coupling phase and an SMG phase for large $y$ where a four fermion condensate forms and $|\sigma^2|$ plateaus.
The corresponding anti-ferromagnetic bilinear and its corresponding susceptibility are shown
in fig.~\ref{kappa0.2}.
The fact that the expectation value of the anti-ferromagnetic bilinear is
independent of $L$ over the region $1.73<y<2.6$ 
suggests a broken phase exists in this region of $y$. The corresponding ferromagnetic bilinear
is close to zero in this window. This
broken phase is bounded by
two phase transitions. We have examined the lattice size $L$ dependence of
the anti-ferromagnetic susceptibility for both of these phase transitions. Indeed, using finite size scaling,
one expects the peak in the susceptibility to scale as a power of $L$ at a phase transition:
\begin{equation}
    \chi_{\rm peak}\sim L^{2-\eta}
\end{equation}
The critical exponent $\eta$ can be found by fitting a straight line to a plot
of $\ln{\chi_{\rm peak}}$ versus $\ln{L}$.
In fig.~\ref{loglog} we show such a plot together with a least squares fit
for the
fermion susceptibility $\chi_{\rm stag}$ at both 
phase transitions. The fits yield $\eta_1 = 0.019\pm 0.077$ for phase transition $L_1$ and $\eta_2 = -0.11\pm 0.15$ for phase transition $L_2$, consistent with $\eta=0$, as expected for mean-field-like bosonic criticality. These values also agree with what was found in the pure four fermion model \cite{Ayyar:2016lxq}. In appendix \ref{moreobs} we show that the width of these peaks can be scaled as $y-y_{1,2}\sim L^{-1.3}$, with $y_{1}(y_2)$ denoting the location of the transition $L_1(L_2)$ at $\kappa=0.2$. \gwen{These critical exponents extracted via finite size scaling of the peak height are consistent with these phase transitions being second-order.}

\begin{figure}
\centering
\includegraphics[width=0.5\linewidth]{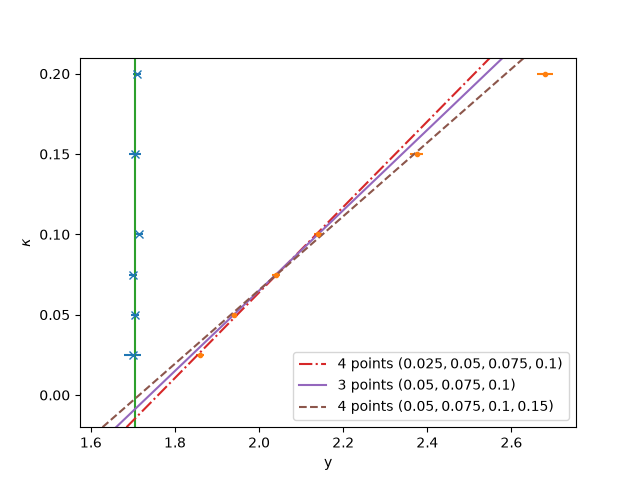}
\caption{Location of the two phase transitions and their crossing point}
\label{extrap}
\end{figure}

By performing a series of sweeps in $y$ at different $\kappa$ we can extrapolate the locations of these
two transitions to their crossing point.
A fit of $L_1$ was found to be vertical, and we also fit subsets of points on $L_2$ at $\kappa = 0.025, 0.05, 0.075, 0.1$ and $0.15$ to extrapolate a point of intersection as show in \fig{extrap}. 
This intersection of fits determines the approximate location of $P_c=(y_c,\kappa_c)$ as occurring at $y_c = 1.706\pm0.003$ and $\kappa_c=-0.01 \pm 0.006$.

\section{Behavior at the merged point $P_c$}
\begin{figure}
    \centering
    \begin{subfigure}[b]{0.48\linewidth}
        \centering
        \includegraphics[width=\linewidth]{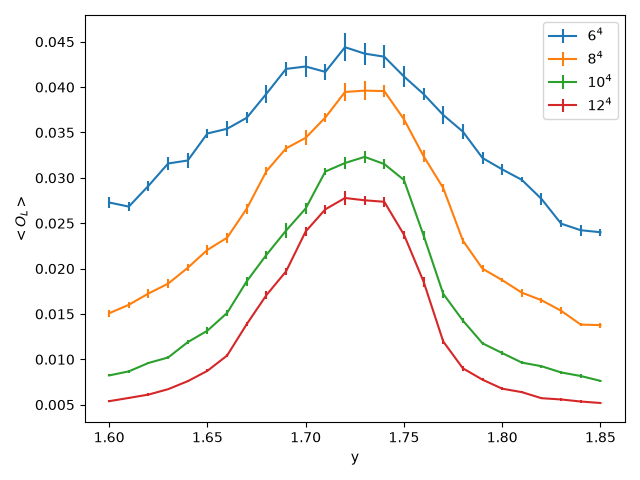}
        \caption{$<O_L>$ vs $y$ at $\kappa=-0.01$}
    \end{subfigure}
    \begin{subfigure}[b]{0.48\linewidth}
        \centering
        \includegraphics[width=\linewidth]{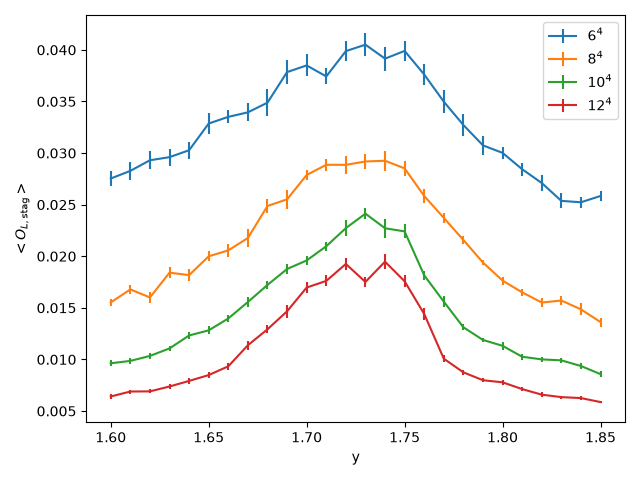}
        \caption{$<O_{L,\rm stag}>$ vs $y$ at $\kappa=-0.01$}
    \end{subfigure}
    \caption{$<O_L>$ and $<O_{L,\rm stag}>$ for $\kappa=-0.01$ and $\lambda=1.0$}
   \label{onelink}
\end{figure}

\begin{figure}
    \centering
    \begin{subfigure}[b]{0.48\linewidth}
        \centering
        \includegraphics[width=\linewidth]{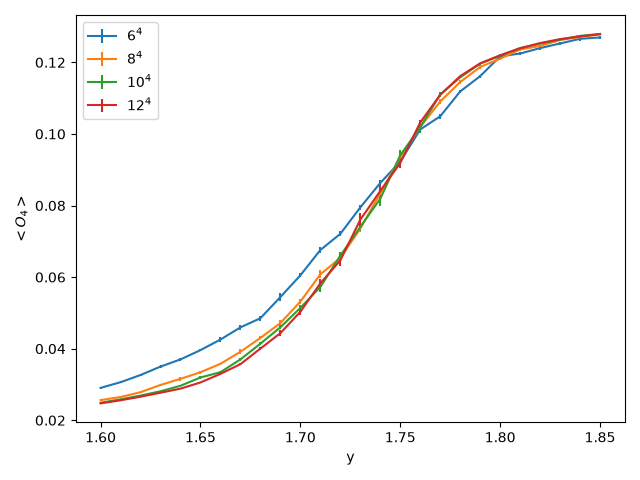}
        \caption{$<O_4>$ vs $y$ at $\kappa=-0.01$}
    \end{subfigure}
    \begin{subfigure}[b]{0.48\linewidth}
        \centering
        \includegraphics[width=\linewidth]{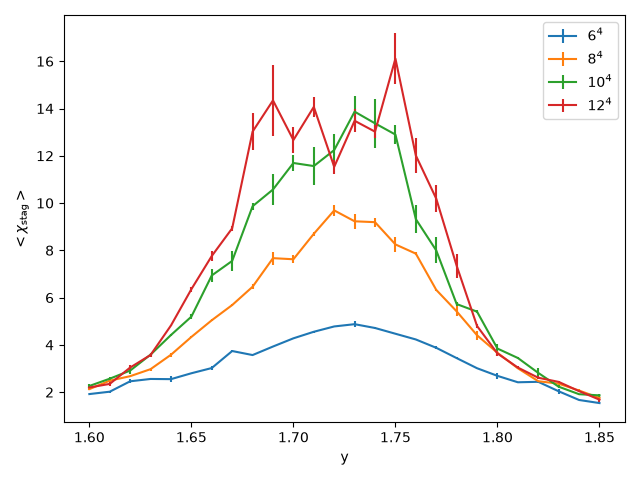}
        \caption{$<\chi_{\rm stag}>$ vs $y$ at $\kappa=-0.01$}
    \end{subfigure}
    \caption{Four fermion condensate $<O_4>$ and $<\chi_{\rm stag}>$ for $\kappa=-0.01$ and $\lambda=1.0$}
    \label{fourkappaneg}
\end{figure}

\begin{figure}
    \centering
    \begin{subfigure}[b]{0.48\linewidth}
        \centering
        \includegraphics[width=\linewidth]{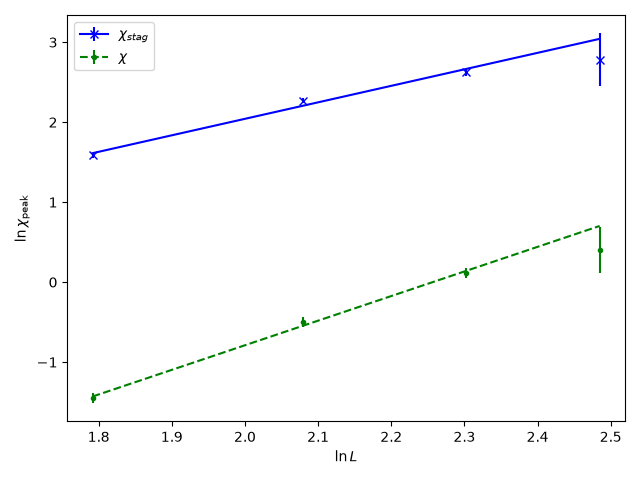}
        \caption{Peak $\chi$ and $\chi_{\rm stag}$ vs $L$ at $\kappa=-0.01$ and $\lambda=1.0$}
    \end{subfigure}
    \begin{subfigure}[b]{0.48\linewidth}
        \centering
        \includegraphics[width=\linewidth]{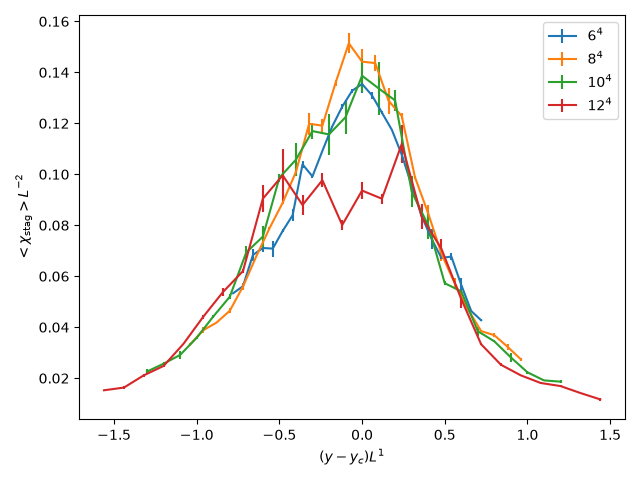}
        \caption{Curve collapse plot for $<\chi_{\rm stag}>$ for $\kappa=-0.01$ and $\lambda=1.0$}
    \end{subfigure}
    \caption{Scaling behavior of $<\chi>$ and $<\chi_{\rm stag}>$ at the critical point $P_c$}
    \label{mergescaling}
\end{figure}

We view the merged point $P_c$ as the most interesting place to look for new critical behavior in the model.
In fig.~\ref{kappaneg} below we show a scan in $y$ of the scalar and fermion bilinears at $\kappa=-0.01$.
\begin{figure}[hb]
    \centering
    \begin{subfigure}[b]{0.48\linewidth}
        \centering
        \includegraphics[width=\linewidth]{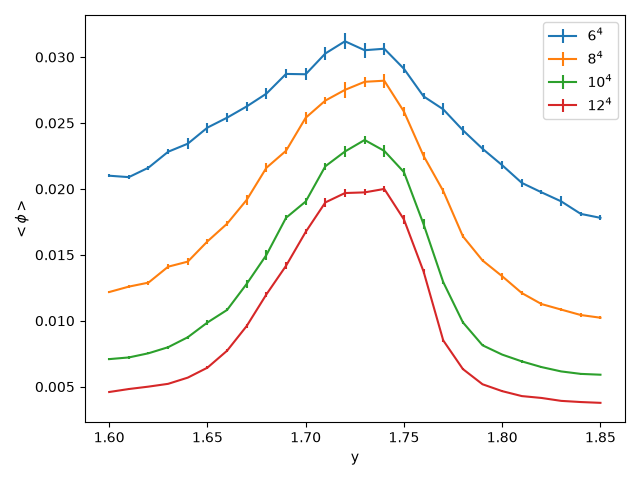}
        \caption{$<\phi>$ vs $y$ at $\kappa=-0.01$}
    \end{subfigure}
    \begin{subfigure}[b]{0.48\linewidth}
        \centering
        \includegraphics[width=\linewidth]{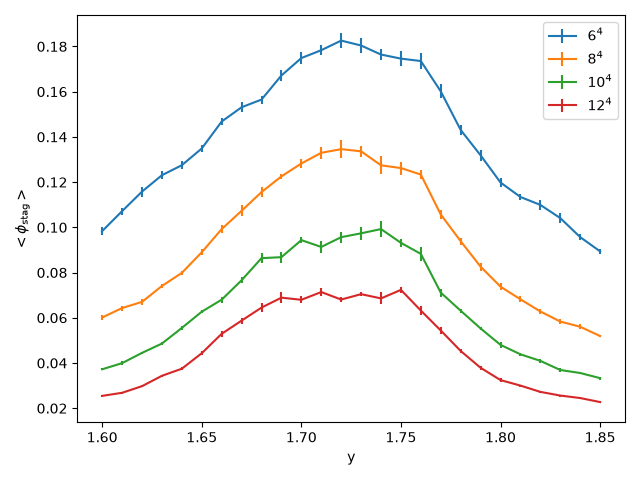}
        \caption{$<\phi_{\rm stag}>$ vs $y$ at $\kappa=-0.01$}
    \end{subfigure}
    \caption{Fermion bilinears vs $y$ at  $\kappa = -0.01$}
    \label{kappaneg}
\end{figure}
It should be clear that both ferromagnetic and anti-ferromagnetic bilinears vanish in the thermodynamic
limit for $\kappa=-0.01$. We have also looked at the one-link bilinear terms corresponding to
\begin{equation}
    O_L=\sum_\mu \xi_\mu(x)\chib(x)\left[\chi(x+\mu)+\chi(x-\mu)\right]
\end{equation}
and
\begin{equation}
    O_{L,\rm stag} = \sum_\mu \epsilon(x)\xi_\mu(x)\chib(x)\left[\chi(x+\mu)+\chi(x-\mu)\right]
\end{equation}
These operators are invariant under the $Z_4$ and $SO(4)$ symmetries but break the shift symmetry.
They are shown in fig.~\ref{onelink} and clearly vanish in the thermodynamic limit for this $\kappa$.
We conclude that at the merged point $P_c$ the expectation values of all fermion bilinears vanish. However, it should also be clear that
this point is not in the weak coupling symmetric phase. Fig.~\ref{fourkappaneg}
shows the four fermion condensate and the anti-ferromagnetic susceptibility as a function of the
Yukawa coupling at $P_c$.
Clearly the system supports a non-zero four fermion condensate at this point. Furthermore, the 
peaks of both the ferro and anti-ferro susceptibilities grow with a non-trivial power of
the lattice size $L$ suggesting that the system is still critical. 
Fitting the power yields the exponents $\eta=-0.06\pm 0.22$ for $\chi_{\rm stag}$ and $\eta^\prime=-1.08\pm 0.15$ as shown in fig.~\ref{mergescaling}. Neither exponent is consistent with the formation of a condensate which agrees with our
direct measurements of the fermion bilinears. In fig.~\ref{mergescaling} we show that the anti-ferromagnetic susceptibility data collapse on to a single curve for different volumes if we assume that the critical region scales  as $y-y_c\sim L^{-1}$. The combined scaling behavior, in particular the strongly enhanced ferromagnetic susceptibility and the $L^{-1}$
 width of the critical region, is not compatible with the mean-field behavior observed on $L_1$ and $L_2$ at positive $\kappa$,
 and suggests unusual critical behavior.
Furthermore, we observe that all the fermion bilinears considered here continue to vanish in the thermodynamic limit as we progress down the critical line for
negative $\kappa$. Indeed, we observe that the susceptibilities still have a peak for some $y=y_c(\kappa)$ when 
$\kappa<0$ (at least for $\kappa>-0.2$) and the height of this peak continues to grow with lattice size.
However, the absolute value of the peak height in $\chi$ decreases as $\kappa$ is made more negative in comparison
to that at the merged point as can be seen in \fig{kappaneg} for
$\kappa=-0.05$.
\begin{figure}
    \centering
    \begin{subfigure}[b]{0.48\linewidth}
        \centering
        \includegraphics[width=\linewidth]{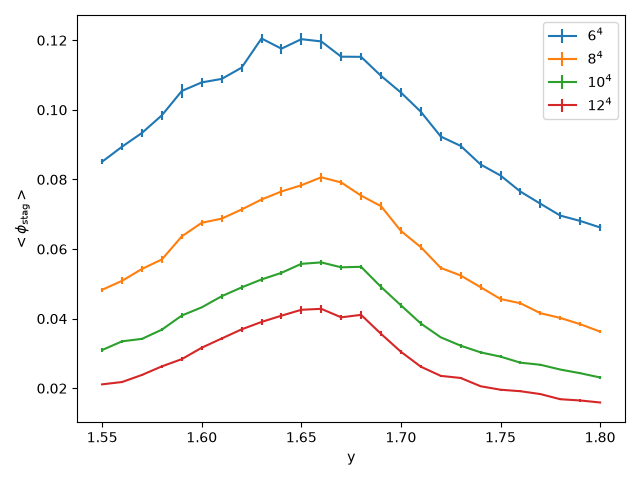}
        \caption{$<\phi_{\rm stag}>$ vs $y$ at $\kappa=-0.05$}
    \end{subfigure}
    \begin{subfigure}[b]{0.48\linewidth}
        \centering
        \includegraphics[width=\linewidth]{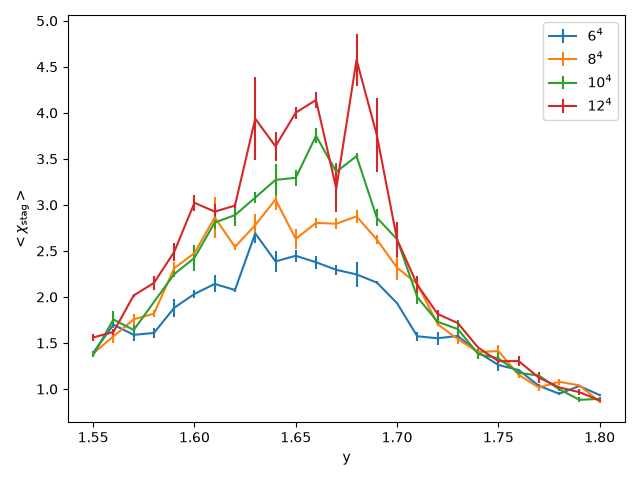}
        \caption{$<\chi_{\rm stag}>$ vs $y$ at $\kappa=-0.05$}
    \end{subfigure}
    \caption{$<\phi_{\rm stag}>$ and $<\chi_{\rm stag}>$ for $\kappa=-0.05$ and $\lambda=1.0$}
    \label{kappaneg}
\end{figure}
This suggests that the correlation length is 
maximal close to or at the merged point $P_c$ consistent with
its identification as a multicritical point.

It is tempting to try and identify the appearance of this multicritical point as corresponding to a lattice effective theory in which the leading kinetic term for $\sigma$ has been tuned to zero. 
If this is so then it would invite a description in terms of 
condensed Hopf defects as in our earlier discussion. However, we have no direct confirmation that
such defects exist in our simulations and so this remain for now  pure speculation which we
hope to address in future work.

\section{Conclusion}
We have explored the phase structure of a model
comprised of two massless staggered fermion fields forming a doublet under a global SU(2), coupled to a scalar in the adjoint
representation of $SU(2)$. In addition to an $SO(4)$ symmetry and shift
symmetries, the model possesses an additional $Z_4$ symmetry which is the surviving subgroup
of the usual $U_\epsilon(1)$ symmetry of staggered fermions in the presence of the Yukawa coupling.
The model is free of sign problems and can be simulated using the RHMC
algorithm. In a two dimensional parameter
space spanned by the Yukawa coupling and the coupling to a scalar kinetic term, we find two
phase boundaries that separate a free massless fermion phase from an intermediate phase of broken $Z_4$
symmetry and an SMG phase at strong coupling.  Using finite-size scaling analysis, we determine the critical exponents associated with the two phase transitions that border the antiferromagnetic phase, obtaining values that are consistent with the predictions of mean-field theory. These phase boundaries come together at a point $P_c$, which we have examined as a potential multicritical point. We provide evidence that a single critical line originates from this merged point and extends into the region of negative $\kappa$.

At $P_c$ we
find evidence that both ferro and antiferromagnetic fermion bilinear order parameters vanish - both the onsite terms and the one link operators. This is strong evidence that
shift symmetry is not broken in the model at this point. Combining these two facts suggests that
the spin-$Z_4$ symmetry (a subgroup of the axial symmetry) is not broken (and likely not anomalous) at $P_c$ which is a key condition for the existence of an SMG phase.
Furthermore, at $P_c$ we observe that the four fermion condensate is non-zero
and fermion susceptibilities diverge with non-trivial critical
exponents. These features persist along the negative $\kappa$ critical line. However the magnitude
of the fermion susceptibility decreases monotonically as we move to more negative $\kappa$. This suggests that the merged point might be a new multicritical point.
We would like to emphasize that
our simulations suffer from extremely long autocorrelation times near $P_c$ 
and it has been very difficult to extract reliable results there in comparison to the phase
transitions observed for positive $\kappa$.

We have tried to interpret this phase diagram from the effective sigma model action that arises
after integration over the fermions. For sufficiently large Yukawa coupling, the effective potential
is minimized on an antiferromagnetic background, while a symmetric gapped phase is naturally
produced as the coupling is sent to infinity. Performing a derivative expansion about this background
produces the usual $O(3)$ sigma model kinetic term and additionally a quartic Skyrme term
that can support topological defects in the continuum. 
These defects are instanton-like configurations that
are classified by the Hopf invariant 
which labels the mappings between the Euclidean spacetime boundary $S^3$
and the $S^2$ vacuum manifold of the scalar theory. In the absence of
the leading kinetic term, these configurations possess a logarithmically divergent action that competes with their logarithmic entropy and can lead to defect condensation in certain regions of the phase
diagram. We conjecture that the merged point corresponds to one such region. However, we
have not tried to measure the topology of the scalar field configurations in this initial study, and so we make no definitive claim for that in this paper.\footnote{ Indeed, we cannot exclude the possibility that these defects, if present, are $Z_2$ defects associated with the mapping $\Pi_4(S^3)$.}.

It is interesting to ask whether one can gauge the $SU(2)$ symmetry.
Rather intriguingly, it appears that Hopf defects remain as
solutions of the equations of motion of a gauged Higgs model with the same
bosonic symmetries \cite{He:2014gqa}. In the background of the Hopf defect, the gauge
symmetry is Higgsed down to $U(1)$. Furthermore, the asymptotic $U(1)$ gauge field 
conspires to force the covariant derivative on the scalar to vanish  
$D\sigma\sim 1/r^2$ as $r\to\infty$. This ensures that
even the scalar kinetic term would give a logarithmic contribution to the
action from such defects and potentially a phase transition to a symmetric condensate for
large Yukawa coupling. We plan to investigate this possibility in the future.
It is possible that the associated phase structure seen in \cite{Butt:2024kxi} may be connected
to such dynamics.

\begin{acknowledgments}
SC and GH were supported by DOE grant DE-SC0009998,  AH by DOE grant DE-SC001005.
We would like to thank Cenke Xu for a careful reading of the paper and useful comments.
\end{acknowledgments}

\bibliographystyle{apsrev4-1} 
\bibliography{smg_staggered_fermions} 
\vfill\newpage

\section{Appendix: Mapping two staggered to four reduced staggered fermions}
\label{map}
The full staggered fermion action we employ in this work is equivalent to the reduced fermion model studied in
\cite{Butt:2018nkn}. To see this transform to the new variables
\begin{align}
    \psi^1(x)&=\frac{1}{\sqrt{2}}(\theta_1(x)+i\theta_2(x))\nonumber\\
    \psib^1(x)&=\frac{1}{\sqrt{2}}(\theta_1(x)-i\theta_2(x))\nonumber\\
    \psi^2(x)&=\frac{1}{\sqrt{2}}(\theta_3(x)+i\theta_4(x))\nonumber\\
    \psib^2(x)&=\frac{1}{\sqrt{2}}(\theta_3(x)-i\theta_4(x))
    \label{transf}
\end{align}
The kinetic term now reads
\begin{equation}
    S_K=\frac{1}{2}\sum_{x,\mu}\sum_{a=1}^4\theta^a(x) \eta_\mu(x)\Delta_\mu \theta^a(x)
\end{equation}
This kinetic term is invariant under an $U(4)$ symmetry corresponding to 
\begin{equation}
    \theta^a(x)\to e^{i\epsilon(x) A}\theta^a(x)
\end{equation}
where $A$ lives in the algebra of $U(4)$.
Substituting the mapping in eqn.~\ref{transf} we find that the Yukawa term $iy\psib \sigma_a\tau_a \psi$ becomes
\begin{align}
    y\left[\sigma_1\left(\theta^2\theta^3+\theta^1\theta^4\right)+\sigma_2\left(\theta^1\theta^3+\theta^4\theta^2\right)+\sigma_3\left(\theta^1\theta^2+\theta^3\theta^4\right)\right]
\end{align}
Introducing the self-dual fermion bilinear
\begin{equation}
    \theta_+^{ab}=\frac{1}{2}(\theta^{ab}+\frac{1}{2}\epsilon^{abcd}\theta^c\theta^d)
\end{equation}
This can be rewritten
\begin{equation}
    2y\theta_a \sigma_+^{ab}\theta_b
\end{equation} 
where we have moved the self-dual projector onto the scalar field. 
In the simple case where the $\sigma$ action contains only a quadratic term ${\rm Tr}(\sigma^2)$ then
integration over $\sigma$
generates a four fermion term of the form $(\bar\psi^a\psi^a)^2$ or the equivalent
reduced term $\epsilon^{abcd}\theta^a\theta^b\theta^c\theta^d$. The presence of the three dimensional
self-dual representation for $\sigma$ indicates that the Yukawa term breaks the $SU(4)$ symmetry to $SO(4)$ while the remaining
$U(1)$ symmetry is broken to $Z_4$.

\section{Appendix: Absence of a sign problem}
\label{sign}
The fermion operator
including both kinetic and Yukawa terms is anti-hermitian. Thus the eigenvalues
lie along the imaginary axis and the determinant is real. Actually because of the
$SU(2)$ symmetry they are also doubly degenerate so it is positive definite.
To see this consider the eigenvalue equation for the fermion operator 
\begin{equation}
    (\eta \cdot\Delta +i\sigma\cdot\tau)\psi=\lambda\psi
\end{equation}
where we can generalize to include a gauged derivative
\begin{equation}
    \Delta_\mu \psi(x)=U_\mu(x)\psi(x+\mu)-U_\mu^\dagger(x-\mu)\psi(x-\mu)
\end{equation}
Taking the complex conjugate we find
\begin{equation}
    (\eta\cdot\Delta^* -i\sigma\cdot\tau^*)\psi^*=\lambda^*\psi^*
\end{equation}
Using $\tau_2 U_\mu \tau_2=U^*$ and pulling $\tau_2$ through the operator from the left leads to
\begin{equation}
    (\eta\cdot\Delta +i\sigma\cdot\tau)\tau_2\psi^*=\lambda^*\tau_2\psi^*
\end{equation}
So every eigenvalue $\lambda$ is paired with another $\lambda^*$ and the
determinant is hence positive semi-definite. Equivalently in the reduced formalism the eigenvalues
come in quartets $(\lambda,\lambda^*,-\lambda,-\lambda^*)$ which renders the Pfaffian real positive
definite as it must.

\section{Appendix: Additional observables and scaling behavior}
\label{moreobs}

\begin{figure}
    \centering
    \begin{subfigure}[b]{0.48\linewidth}
        \centering
        \includegraphics[width=\linewidth]{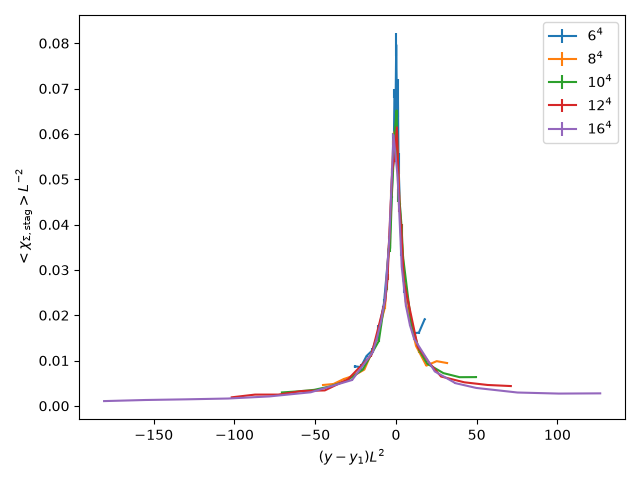}
        \caption{Transition $L_1$ from symmetric to antiferromagnetic phases}
    \end{subfigure}
    \begin{subfigure}[b]{0.48\linewidth}
        \centering
        \includegraphics[width=\linewidth]{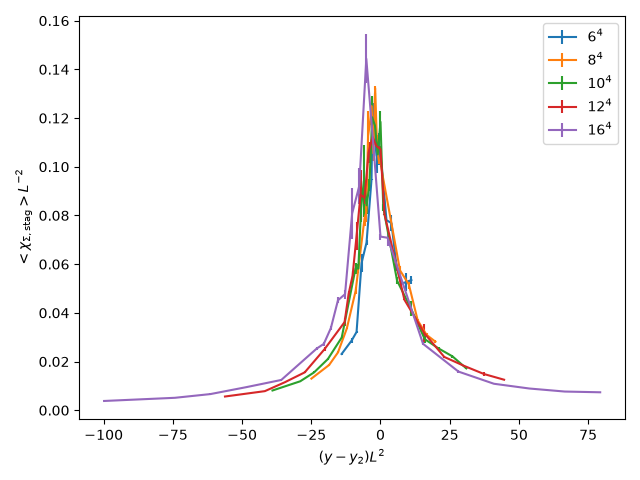}
        \caption{Transition $L_2$ from antiferromagnetic to SMG phases}
    \end{subfigure}
    \caption{Curve collapse plots for $<\chi_{\Sigma,\rm stag}>$ for $\kappa=0.2$ and $\lambda=1.0$}
    \label{scalingsigma}
\end{figure}

\begin{figure}
    \centering
    \begin{subfigure}[b]{0.48\linewidth}
        \centering
        \includegraphics[width=\linewidth]{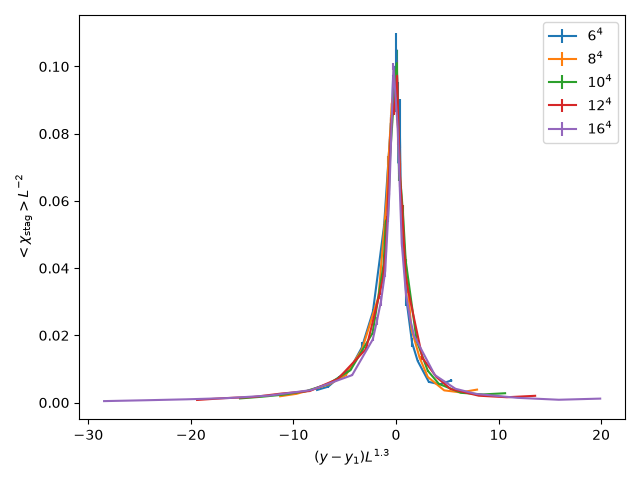}
        \caption{Transition $L_1$ from symmetric to antiferromagnetic phases. }
    \end{subfigure}
    \begin{subfigure}[b]{0.48\linewidth}
        \centering
        \includegraphics[width=\linewidth]{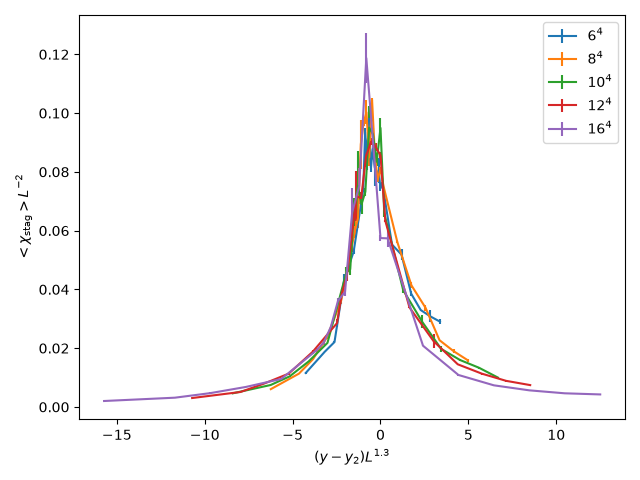}
        \caption{Transition $L_2$ from antiferromagnetic to SMG phases}
    \end{subfigure}
    \caption{Curve collapse plots for $<\chi_{\rm stag}>$ for $\kappa=0.2$ and $\lambda=1.0$}
    \label{scalingbilin}
\end{figure}

We have also looked at the ferromagnetic and antiferromagnetic susceptibility of the scalar field given by
\begin{equation}
    \chi_{\Sigma}=V(<\Sigma^2>-<\Sigma>^2)\quad\text{and}\quad\chi_{\Sigma,\rm stag}=V(<\Sigma_{\rm stag}^2>-<\Sigma_{\rm stag}>^2)
\end{equation}
While both $<\chi_{\rm stag}>$ and $<\chi_{\Sigma,\rm stag}>$ are consistent with $\eta = 0$ along both transitions $L_1$ and $L_2$ at $\kappa=0.2$ and $\lambda=1.0$, they show different scaling in the width of the peaks. In figure \ref{scalingsigma} we show that the scalar susceptibility data collapse onto a single curve for all volumes if we assume that the critical regions have $y-y_{1,2}\sim L^{-2}$, where $y_1=1.706$ and $y_2=2.69$, which is consistent with bosonic mean field theory. The fermion susceptibility data however only collapses with a different scaling of $y-y_{1,2}\sim L^{-1.3}$ instead, which can be seen in figure \ref{scalingbilin}.

\end{document}